\documentclass{camera}
\usepackage{graphicx}  % uncomment this if your want to insert figures

\begin{document}
\begin{flushright}
UPRF-2003-30\\
LPTHE-P03-24\\
\vspace{-12pt}
\end{flushright}
%%%%%%%%%%%%%%%%%%%%%%%%%%%%%%%%%%%%%%%%%%%%%%%%%%%%%%%%
% The title, all uppercase; if you want to split it in
% two or more lines, put a \\ macro at each line break
% example: 
%   \title{TITLE: FIRST LINE\\ SECOND LINE}
%
\title{HEAVY QUARK PRODUCTION: THEORY VS. EXPERIMENT
$^\clubsuit$
}
\footnotetext{$\clubsuit$ Talk given
at IFAE Lecce (Italy), April 2003}
%%%%%%%%%%%%%%%%%%%%%%%%%%%%%%%%%%%%%%%%%%%%%%%%%%%%%%%%
% The author(s), separated by commas; do not put a
% comma before the last author, use instead the \and
% macro which produces a normal ``and'' in the
% caps/small caps context
%
\author{Matteo Cacciari}

%%%%%%%%%%%%%%%%%%%%%%%%%%%%%%%%%%%%%%%%%%%%%%%%%%%%%%%%
%
\organization{Dipartimento di Fisica, Universit\`a di Parma,\\
INFN, Sezione di Milano, Gruppo Collegato di Parma, Italy, and\\
LPTHE, Universit\'e Paris 6, France}

\maketitle

%%%%%%%%%%%%%%%%%%%%%%%%%%%%%%%%%%%%%%%%%%%%%%%%%%%%%%%%
% Write the text starting from here and using the usual
% LaTeX commands.
%

Heavy quark production has been for years a closely scrutinized process, as  it
has represented  one of the very few instances in which  experimental
measurements and next-to-leading order (NLO) QCD predictions seemed to be at
variance. Such a disagreement appeared surprising, as one expects perturbative
QCD to be able to handle heavy quark predictions well, due to the (relatively) 
large scale set by their mass and to their small hadronization
corrections\footnote{It should however be noted that, while the large heavy
quark mass does indeed make accurate calculations possible,  at the same time
it raises the bar: heavy quark production rates  become calculable and they are
a real {\sl prediction} of QCD, contrary to the  light quark case. Recalling
that the NLO corrections  predict a large increase with respect to the leading
order rates, it should not come as a surprise if this calculation does not yet
manage to describe perfectly the measurements.}. Recently, new theoretical
developments and better use of higher-quality   non-perturbative information
have however  greatly reduced the disagreement in bottom production in $p\bar
p$ collisions,  to the point that it does not appear significant
anymore~\footnote{Recent comparisons~\cite{Cacciari:2003uh} to new preliminary
bottom data from the Tevatron Run II show an even better  agreement. I am not
reviewing here $ep$ and $\gamma\gamma$ collisions, where discrepancies still
seem to exist, but where the accuracy of both data and theoretical predictions
is also lower.}. At the same time,  comparisons for top and charm production
appear successful, pointing once again to a healthy status for this sector of
QCD.\\

\begin{figure}
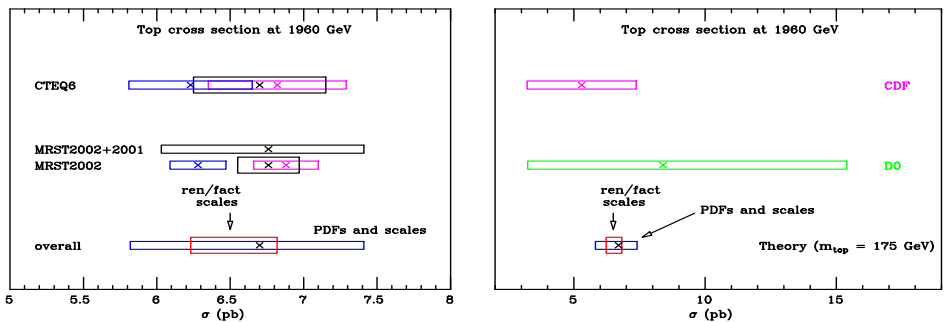

\begin{center}
\includegraphics[height=4.2cm,width=6cm]{top-theory.seps}
~~
\includegraphics[height=4.2cm,width=6cm]{top.seps}
\vspace{-.4truecm}
\caption{Left: theoretical predictions and uncertainties for the
top total cross section. Right: comparison with CDF and D0 preliminary
measurements available in April 2003.}
\label{fig1} 
\end{center}
\end{figure}

\noindent
{\Large\bf Top -}
The top quark was discovered at the Tevatron during Run I, and it has
recently been once again observed in the first Run II data. Its total
cross section, now measured at the slightly higher centre-of-mass
energy $\sqrt{S}=1960$~GeV, can therefore be compared to new
predictions. While NLO corrections~\cite{Nason:1987xz} 
for the total cross section were
already available at the time of the top discovery, improvements of the
last few years include the implementation of the resummation of
soft-gluon  (threshold) effects to next-to-leading logarithmic (NLL)
accuracy~\cite{Bonciani:1998vc}, 
and the determination of parton distribution function sets
(PDFs) with associated errors~\cite{Pumplin:2002vw}.
% This last tool, in particular, allows
% one to try to assess the uncertainty due to the  PDFs with a possibly
% higher degree of rigour and confidence then previously possible, when
% one could only rely on comparing the predictions of different PDF sets.
% In fact, this was  of course an inherently unsatisfactory procedure, as
% the sets themselves  are fitted to the very same experimental
% data and their possible differences do not faithfully represent  the
% degree of our ignorance.
Figure~\ref{fig1} displays in the left panel the theoretical predictions
and corresponding uncertainties~\cite{Cacciari:2003fi} for the top total 
cross section at the
Tevatron Run II, i.e. $p\bar p$ collisions at $\sqrt{S}=1960$ GeV. 
For each PDF set the crosses correspond to the prediction obtained by
setting the renormalization and factorization equal to 1/2, 1 or 2 times
the central scale. The rectangle around each
cross represents instead the uncertainty related to the given PDF set.
% , in each
% case determined according to $\Delta {\cal O} = \frac{1}{2}  
% \sqrt {\sum_{i=1,n_{PDF}} \, ({\cal
% O}_{i+} - {\cal O}_{i-} )^2  }$, where the sum runs on all the sets
% which, within a given PDF determination, represent an ``allowed''
% variation of each of the fit parameters. It is believed that such a
% procedure allows for a sensible determination of the associated error
% on a given observable. In fact, by looking at the plot one notices that
% this is probably not yet entirely true, as the errors predicted by different
% sets do not always overlap. In the right pane the overall theoretical
% prediction with its uncertainties is compared to the measurements
% presently available.
It is apparent from the right panel of figure~\ref{fig1} that the experimental 
uncertainties are nowadays still much larger than the theoretical ones, and
that the latter are dominated by the PDFs uncertainty, due to our poor
knowledge of the gluon distribution function at large $x$. This points to a
situation where not a  better calculation, but rather more accurate
experimental inputs will be needed in order to perform a more compelling
comparison for this  observable.\\

\noindent
{\Large\bf Bottom -}
Bottom production in $p\bar p$ collisions at UA1 and the
Tevatron~\cite{Albajar:1987iu} has
been for years an example of a possible discrepancy between
theory and data, by factors of 2 to 3.  Many explanations for the
discrepancy were proposed, from conventional ones (like the importance of
higher orders or small-$x$ contributions) to a more exotic one involving
supersymmetric particles production~\cite{Berger:2000mp}.

A  careful analysis of the way the data are extracted and/or theoretical
predictions are evaluated shows however that some common
practices need to be revised in the light of today's desired accuracy.
Bottom quarks are not observed as free quarks, as they hadronize into $H_b$
hadrons before decaying to other bottomless final states. The
$b\to H_b$ transition cannot be described by perturbative QCD, and it is
usually parameterized by convolving the momentum distribution of the quark
with a phenomenological function - usually extracted from $e^+e^-$ data
- 
which accounts for the degradation of its momentum in the hadronization process. 
%Over the years, and according
%to
%fits to $e^+e^-$ data, it had become common practice to use the
%Peterson et al. functional form, with its free parameter $\epsilon$
%usually taken in the range $0.002-0.006$.
%This practice overlooks however a fundamental fact: 
The bottom quark
not being directly observed, there is however no way to determine uniquely  its
`fragmentation': when measuring the $H_b$ cross section we only 
observe the result of both a perturbative (hard gluon emissions) and a
non-perturbative (soft gluons) degradation of the initial quark
momentum. Hence, in the theoretical description 
the two `steps' must be properly matched, and no  single function can
be a tool good for all purposes and instances. 
%Even varying the parameter
%of the Peterson function within the above range does in no way guarantee
%that one is indeed exploring the full uncertainty range.
Data  presented at the unphysical $b$
quark level, the result of a deconvolution performed on real $H_b$ hadron
data, may therefore be tainted by the use of an improper phenomenological
parameterization of the fragmentation effects. Fortunately, the CDF
Collaboration has also published data for the $B^+$
mesons, real observable objects, allowing for a safer comparison
between theory and experiment.

\begin{figure}
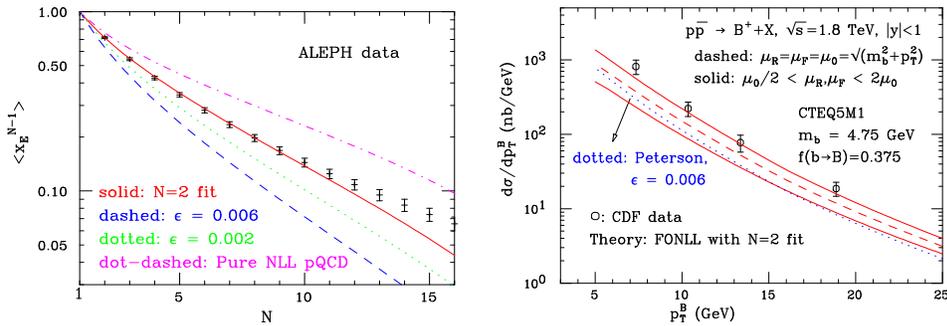

%\begin{center}
\includegraphics[height=4.2cm,width=6cm]{cfr-kart-pet.seps}
~~
\includegraphics[height=4.2cm,width=6cm]{B-hadrons.seps}
\vspace{-.4truecm}
\caption{Left: Fit to $e^+e^-$ moment space data. 
Right: The new prediction for $B^+$ production at the Tevatron,
compared to the experimental data}
\label{fig2} 
%\end{center}
\end{figure}

% A related - but distinct - issue, is what kind of non-perturbative
% information is most important to describe $B$ transverse momentum
% distributions at the Tevatron. It was noted a few years ago that, being
% the $b$ quark $p_T$ spectrum fairly steep and falling with a fairly
% constant power $\sim A/p_T^5$, only the Mellin moments of the
% non-perturbative function around $N=5$ are important in determining the
% $B$ meson cross section.

Ref.~\cite{Cacciari:2002pa} implements these considerations by 
properly matching
perturbative 
(FONLL, i.e. full massive fixed order calculation to NLO accuracy
plus resummation of $\log(p_T/m)$ terms to NLL accuracy) 
and non-perturbative
physics, extracting the relevant (i.e. moments around $N=5$) experimental input 
from $e^+e^-$ data employing the same kind of perturbative calculation
(and the same parameters) which will then be used to calculate the cross
section at the Tevatron.  The results of this analysis are shown in
fig.~\ref{fig2}. The data and the theoretical predictions are
compatible within the uncertainties.

\begin{figure}
%\begin{center}
\includegraphics[height=4.2cm,width=12.5cm]{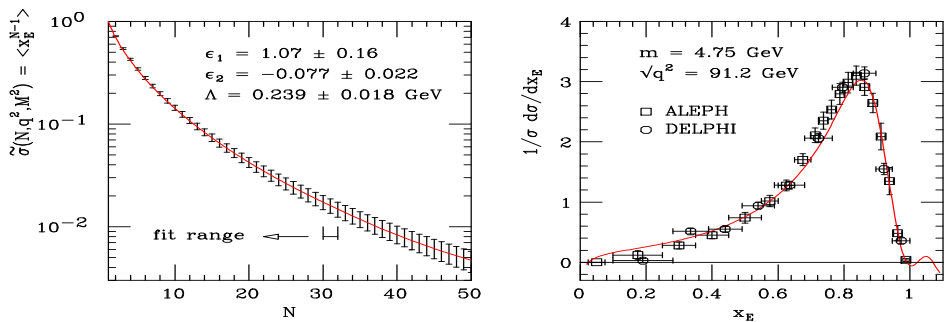}
\vspace{-.4truecm}
\caption{Left: Fit to $e^+e^-$ moment space data. 
Right: The same curve in $x$-space}
\label{fig3} 
%\end{center}
\end{figure}

More theoretical progress in the bottom production sector has taken place for 
the heavy quark fragmentation function. Large-$N$ moments are not needed to
describe hadroproduction, but they are well measured in $e^+e^-$ collisions,
and have an enhanced sensitivity to non-perturbative contributions. All
finite order perturbative calculations, and even finite logarithmic
accuracy resummations, are bound to fail eventually when the endpoint
region is approached. An `all-logs' calculation can however be
performed in the large-$\beta_0$ limit~\cite{Cacciari:2002xb}. Including 
infinitely many orders in the strong coupling this so-called Dressed Gluon
Exponentiation (DGE) result is necessarily divergent,
as perturbative QCD series are only asymptotic. 
The divergence manifests itself in terms of poles (``renormalons'') 
in the Borel transform, which becomes non-invertible unless a
regularization prescription is supplied.
% In this case a
% Principal Value prescription was adopted   
% at each of the poles while inverting the Borel transform of the series. 
The
ambiguity of the regularization procedure is related to the missing higher
twist terms, whose functional form can therefore now be inferred. Hence
one is left with a perturbative calculation and with the matching power
corrections which build up its complementary non-perturbative function. 
The full result can be written as a convolution of the two terms, 
$D(x) = D^{PT}(x) \otimes D^{NP} (x)$, neither independent of the other.
Being the functional form for $D^{NP}$ suggested by the ambiguity, 
there is no need to resort to a phenomenological model, and
only a few parameters have to be fixed by using the data. The fragmentation
function $D(x)$ resums effects at the scale $m/N$, and it is
furthermore universal. It can be convoluted with a specific coefficient
function in order to describe a given process, e.g. $e^+e^-$.
Figure~\ref{fig3} shows how well the data can be described just by fitting
 the non-perturbative function
\begin{equation}
\small
D_N^{NP}\left({\epsilon_1},{\epsilon_2},\frac{(N-1){\Lambda_{QCD}}}{m}\right) =
\exp\left\{-{\epsilon_1} \frac{(N-1){\Lambda_{QCD}}}{m} 
- {\epsilon_2} \left(\frac{(N-1){\Lambda_{QCD}}}{m}\right)^2\right\} 
\label{NP}
\end{equation}
to a very limited set of moments. The leading non-perturbative
effect, a shift to the left of the perturbative distribution controlled
by $\epsilon_1\Lambda_{QCD}$, is compatible with the expectation that this 
be a typical hadronic scale, i.e. $\sim 300$ MeV. It
is also worth mentioning how, fitting also $\Lambda_{QCD}$ in order to check
for consistency, this  does indeed return a value fully compatible with other
determinations of the strong coupling.\\[-1pt]

\noindent
{\Large\bf Charm -}
A QCD calculation for bottom production should also be able to reproduce charm
data, given the replacement $m_b\to m_c$ and a different
non-perturbative part, since the quark $\to$ meson transition is
quantitatively different.

\begin{figure}
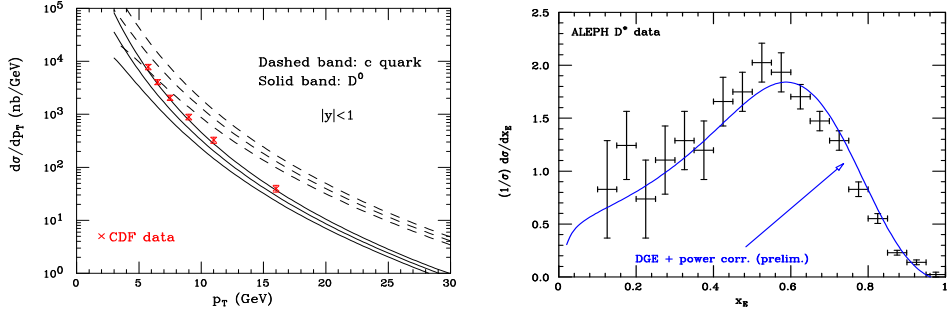

%\begin{center}
\includegraphics[height=4.1cm,width=6cm]{D0.seps}
~~
\includegraphics[height=4.cm,width=6cm]{c-aleph.seps}
\vspace{-.3truecm}
\caption{Left: Theory and data for $D^0$ production in proton-proton collisions
at the Tevatron Run II. Right: description of the ALEPH $D^*$ fragmentation data at
LEP with the DGE calculation and the non-perturbative function fitted
in~\protect\cite{Cacciari:2002xb} to $B$
fragmentation, with only the $m_b\to m_c$ replacement~\protect\cite{CGprelim}.}
\label{fig4} 
%\end{center}
\end{figure}

To verify this, the same theoretical framework which gives a
fairly good description of the $B^+$ CDF data at the Tevatron has been checked
against new charm data from the Tevatron Run II~\cite{Acosta:2003ax}. 
In this case~\cite{Cacciari:2003zu} the determination
of the non-perturbative function was slightly more involved, as LEP data were
not available for all the states ($D^0$, $D^+$, $D^*$ and $D_s$) measured in
hadronic collisions. Effective fragmentation functions were therefore
constructed from the one for $D^*$, which is accurately measured in
$e^+e^-$, modeling the $V\to P$ decays and  employing a perturbative QCD 
model~\cite{Braaten:1995bz} 
to parameterize fragmentation into vector
($V$) and pseudoscalar ($P$) mesons with one free parameter only.
The results for $D^0$ are shown in figure~\ref{fig4} (those for $D^+$ and
$D^{*+}$ are qualitatively identical).  The FONLL calculation, complemented by
non-perturbative information properly extracted from $e^+e^-$ 
data, is clearly able to describe the CDF data within
the uncertainties.

Last but not least, the resummed calculation of
ref.~\cite{Cacciari:2002xb} and the form it predicts for the
non-perturbative power corrections, given in eq.~(\ref{NP}), can be
tested against $D^*$ fragmentation data in $e^+e^-$ collisions.
Figure~\ref{fig4} shows in the right panel the ALEPH data and 
the curve given by the same non-perturbative parameters
fitted to $B$ meson data and listed in fig.~\ref{fig3}, the only change
having been the replacement of the
bottom mass by the charm  mass in both the perturbative and the
non-perturbative components of the fragmentation function. The fairly
good description of the data, at this  stage to be taken only as a
preliminary indication~\cite{CGprelim}, shows that the scaling of the
non-perturbative effects with the heavy quark mass is indeed correctly
predicted, and that from the numerical point of view the hadronization
effects are pretty similar in the $B$ and the $D$ sector, though a more
refined analysis should certainly be performed.

\end{document}